\documentclass{midl} % Include author names
\usepackage{mwe} % to get dummy images
\usepackage{multirow}

\jmlrproceedings{MIDL 2020}{Medical Imaging with Deep Learning 2020}
\jmlrvolume{}
\jmlryear{}
\jmlrworkshop{MIDL 2020 -- Short Paper}
\editors{}
\title[BraTS Uncertainty Evaluation]{Uncertainty Evaluation Metric for Brain Tumour Segmentation}

\midlauthor{\Name{Raghav Mehta} \Email{raghav@cim.mcgill.ca}\\
\addr Centre for Intelligent Machines, McGill University, Canada
\AND
\Name{Angelos Filos} \Email{angelos.filos@cs.ox.ac.uk}\\
\Name{Yarin Gal} \Email{yarin.gal@cs.ox.ac.uk} \\
\addr Oxford Applied and Theoretical Machine Learning Group, University of Oxford, England
\AND
\Name{Tal Arbel} \Email{arbel@cim.mcgill.ca}\\
\addr Centre for Intelligent Machines, McGill University, Canada
}

\begin{document}

\maketitle

\begin{abstract}
In this paper, we develop a metric designed to assess and rank uncertainty measures for the task of brain tumour sub-tissue segmentation in the BraTS 2019 sub-challenge on uncertainty quantification. The metric is designed to: (1) reward uncertainty measures where high confidence is assigned to correct assertions, and where incorrect assertions are assigned low confidence and (2) penalize measures that have higher percentages of under-confident correct assertions. Here, the workings of the components of the metric are explored based on a number of popular uncertainty measures evaluated on the BraTS 2019 dataset. 

\end{abstract}


\section{Introduction}
Deep Neural Networks (DNN) have shown to outperform traditional machine learning methods on a variety of automatic medical image segmentation tasks \cite{nn-Unet,NeoNatal,DeepOrgan}, including tumour segmentation, as depicted by the highest ranking results on recent BraTS challenges~\cite{BraTS2018}. However, errors in brain tumour segmentation deter the adoption of DNN frameworks in clinical contexts, particularly those that rely on high voxel-level accuracy, such as in image-guide neurosurgery.  Although popular DNNs~\cite{3DUNet,DeepMedic} for brain tumour segmentation provide the "sigmoid"/"softmax" predictions for tumour labels, it is the overall {\it model uncertainties} which would more informative in assisting clinicians in making more informed decisions. Although, several recent methods~\cite{MCD,UEDE,SWAG} have been proposed to estimate uncertainties in deep neural networks, there is no established strategy in which their usefulness can be assessed and compared for particular clinical contexts. In this paper, we develop metrics to measure the quality of different uncertainty measures for the task of brain tumour segmentation, with the objectives: (1) when the network is correct it is confident in the predicted labels, and (2) when they are incorrect, it is not confident. These metrics were combined and used as the basis of ranking the uncertainties produced by participating teams in the BraTS 2019 sub-challenge on uncertainty quantification. 
%It should be noted that this metric is designed assuming we do not have access to the "ground truth" uncertainty of inter-rater variability, which requires multiple annotations.

\section{Metric for Assessing and Comparing Uncertainty Measures}

In the context of the BraTS 2019 challenge, each team provided their multi-class brain tumour segmentation output labels and the voxel-wise uncertainties for each of the associated tasks: whole tumour (WT), tumour core (TC) and enhanced tumour (ET) segmentations. For each task, the uncertain voxels were filtered out at several predetermined (N) number of uncertainty thresholds, $\tau$, and the model performance was assessed based on the contextual metric of interest (here, Dice score) on the remaining voxels at each of the thresholds. For example, at $\tau = 0.75$, all voxels with uncertainty values $\ge 0.75$ are marked as uncertain. The associated predictions are filtered out, and Dice values are calculated for the remaining predictions based on the unfiltered voxels. This evaluation rewards models where the confidence in the incorrect assertions (False Positives - FPs, and False Negatives - FNs) is low and high for correct assertions (True Positives - TPs and True Negatives - TNs). For these models, it is expected that as more uncertain voxels are filtered out, the Dice score should increase on the remaining predictions.

\begin{figure*}[t]
\centering
\includegraphics[width=0.85\textwidth]{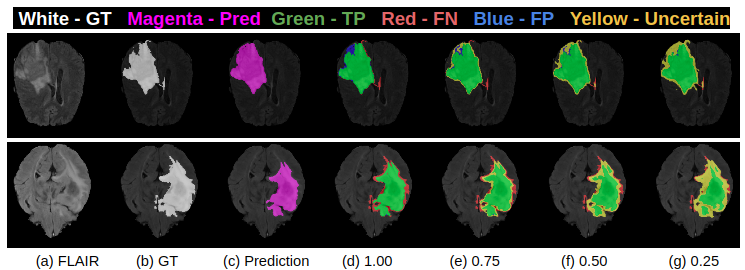}\vspace{-3mm}
\caption{Effect of uncertainty thresholding on  two examples for whole tumour segmentations (Top and bottom rows). (a) FLAIR MRI, (b) "Ground truth" labels, (c) Sample prediction, (d) Prediction with no filtering, and  (e)-(g) Filtering with uncertainty thresholds ($\tau$) of 0.75, 0.5 and 0.25.}
\label{fig:UncExam}
% \vspace{-3mm}
\end{figure*}

\begin{table}[t]
\scriptsize
% \resizebox{\textwidth}{!}{
\centering
\begin{tabular}{|l|l|l|l|l|}
\hline
\multirow{2}{*}{}  & \multicolumn{4}{c|}{\textbf{Dice}}                                                            \\ \cline{2-5} 
                   & \textbf{Dice at 1.00} & \textbf{Dice at 0.75} & \textbf{Dice at 0.50} & \textbf{Dice at 0.25} \\ \hline
\textbf{Example-1} & 0.94                  & 0.96                  & 0.965                 & 0.97                  \\ \hline
\textbf{Example-2} & 0.92                  & 0.955                 & 0.97                  & 0.975                 \\ \hline
\multirow{2}{*}{}  & \multicolumn{4}{c|}{\textbf{Ratio of Filtered TPs (($\text{TP}_{\text{1.00}}$ - $\text{TP}_\tau$) / $\text{TP}_{\text{1.00}}$)}}                \\ \cline{2-5} 
                   & \textbf{FTP at 1.00}  & \textbf{FTP at 0.75}  & \textbf{FTP at 0.50}  & \textbf{FTP at 0.25}  \\ \hline
\textbf{Example-1} & 0.00                  & 0.00                  & 0.05                  & 0.1                   \\ \hline
\textbf{Example-2} & 0.00                  & 0.00                  & 0.15                  & 0.25                  \\ \hline
\multirow{2}{*}{}  & \multicolumn{4}{c|}{\textbf{Ratio of Filtered TNs (($\text{TN}_{\text{1.00}}$ - $\text{TN}_\tau$) / $\text{TN}_{\text{1.00}}$)}}                \\ \cline{2-5} 
                   & \textbf{FTN at 1.00}  & \textbf{FTN at 0.75}  & \textbf{FTN at 0.50}  & \textbf{FTN at 0.25}  \\ \hline
\textbf{Example-1} & 0.00                  & 0.0015                & 0.0016                & 0.0019                \\ \hline
\textbf{Example-2} & 0.00                  & 0.0015                & 0.0026                & 0.0096                \\ \hline
\end{tabular}%}
\caption{Changes in Dice, Filtered True Positives (FTP), and Filtered True Negatives (FTN) with different uncertainty thresholds ($\tau$) for two different examples.}
\label{tab:unc_exam}
% \vspace{-5mm}
\end{table}

\begin{figure*}[t]
\centering
\includegraphics[width=0.9\textwidth]{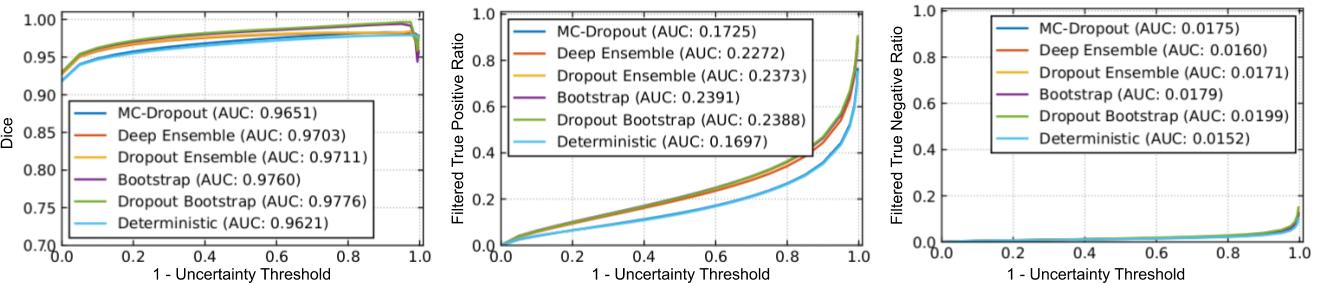} \vspace{-3mm}
\center \caption{Effect of changing uncertainty threshold ($\tau$) on whole tumour for entropy measure.}
\label{fig:UncEntropy}
% \vspace{-5mm}
\end{figure*}

The proposed strategy does not keep track of the number of correctly labeled voxels that are filtered at each threshold level along with the uncertain incorrect labels. In order to penalize filtering out many correctly predicted voxels (TPs, TNs) when attaining high Dice values, an additional assessment component is added to keep track of the filtered TP and TNs voxels. Given that tumour segmentation is expected to have a high-class imbalance between tumour and healthy tissues, the system keeps track of the filtered TPs and TNs separately. 
The ratio of filtered TPs (FTP) at different thresholds ($\tau$) is measured relative to the unfiltered values ($\tau = 1.00$) such that FTP =  ($\text{TP}_{\text{1.00}}$ - $\text{TP}_\tau$) / $\text{TP}_{\text{1.00}}$. The ratio of filtered TNs is calculated in a similar manner. This evaluation essentially penalizes approaches that filter out a large percentage of TP or TN relative to $\tau=1.00$ voxels in order to attain the reported Dice value. 

Figure~\ref{fig:UncExam} and Table~\ref{tab:unc_exam} shows the workings of the assessment metric for example cases based on images from BraTS 2019. Decreasing the threshold ($\tau$) leads to filtering out voxels with incorrect assertions,  leading to an increase in the Dice value for the remaining voxels. Case 2 shows a marginally higher Dice value than Case 1 at uncertainty thresholds $\tau$ = 0.50 and 0.25. However, the Ratio of Filtered TPs and TNs indicates that this is at the expense of marking more TPs and TNs as uncertain.

Finally, different uncertainty measures are ranked according to a unified score which combines the area under three curves: 1) Dice vs $\tau$, 2) FTP vs $\tau$, and 3) FTN vs $\tau$, for different values of $\tau$. The unified score is calculated as follows: 
\begin{equation}
score = \frac{{AUC}_1 + (1-{AUC}_2) + (1-{AUC}_3)}{3}.
\label{Equation-score}
\end{equation}

% \vspace{-3mm}
\section{Experiments and Results}
A modified 3D U-Net architecture~\cite{3DUNet,mehta20183d} generates the segmentation outputs and corresponding uncertainties. We train (228), validate (57), and test (50) this network based on the publicly available Brain Tumour Segmentation (BraTS) challenge 2019 training dataset (335)~\cite{BraTS2018}. The performances of whole tumour segmentation with the Entropy uncertainty measure~\cite{UNAL}, which captures the average amount of information contained in the predictive distribution, using MC-Dropout \cite{MCD}, Deep Ensemble \cite{UEDE}, Dropout Ensemble \cite{EMCD}, Bootstrap, Dropout Bootstrap, and Deterministic, are  shown in Figure~\ref{fig:UncEntropy}. \footnote{Please refer to supplementary material for more results.} Dropout bootstrap shows the best Dice performance (highest AUC), but also has the worst performance for Filtered True Positive and Filtered True Negative curves (highest AUC). This result shows that, here, the higher performance in Dice is at the expense of a higher number of filtered correct voxels. Overall, the metric is working in line with the objectives. However, there is no clear winner amongst these uncertainty methods in terms of rankings. 
 
% \vspace{-3mm} 
\section{Conclusion}
This paper provides a rationale behind the design of a metric presented at the MICCAI BraTS 2019 sub-challenge  to evaluate and rank uncertainties produced by different methods for brain tumour segmentation. Using two different examples it was demonstrated that the designed metric is able to reward methods which convey higher uncertainty for incorrect assertions and penalize methods which have higher uncertainties for correct assertions. 
%This shows the usefulness of the designed metric and encourages to analyze it further. 

\bibliography{midl-samplebibliography}

\newpage
\section*{Supplementary Material}
\beginsupplement

%%%%%%%%%%%%%%%%%%%%%%%%%%%%%%%%%%%%%%%%%%%%%%%%%
\begin{figure*}[h!]
\centering
\includegraphics[width=1.00\textwidth]{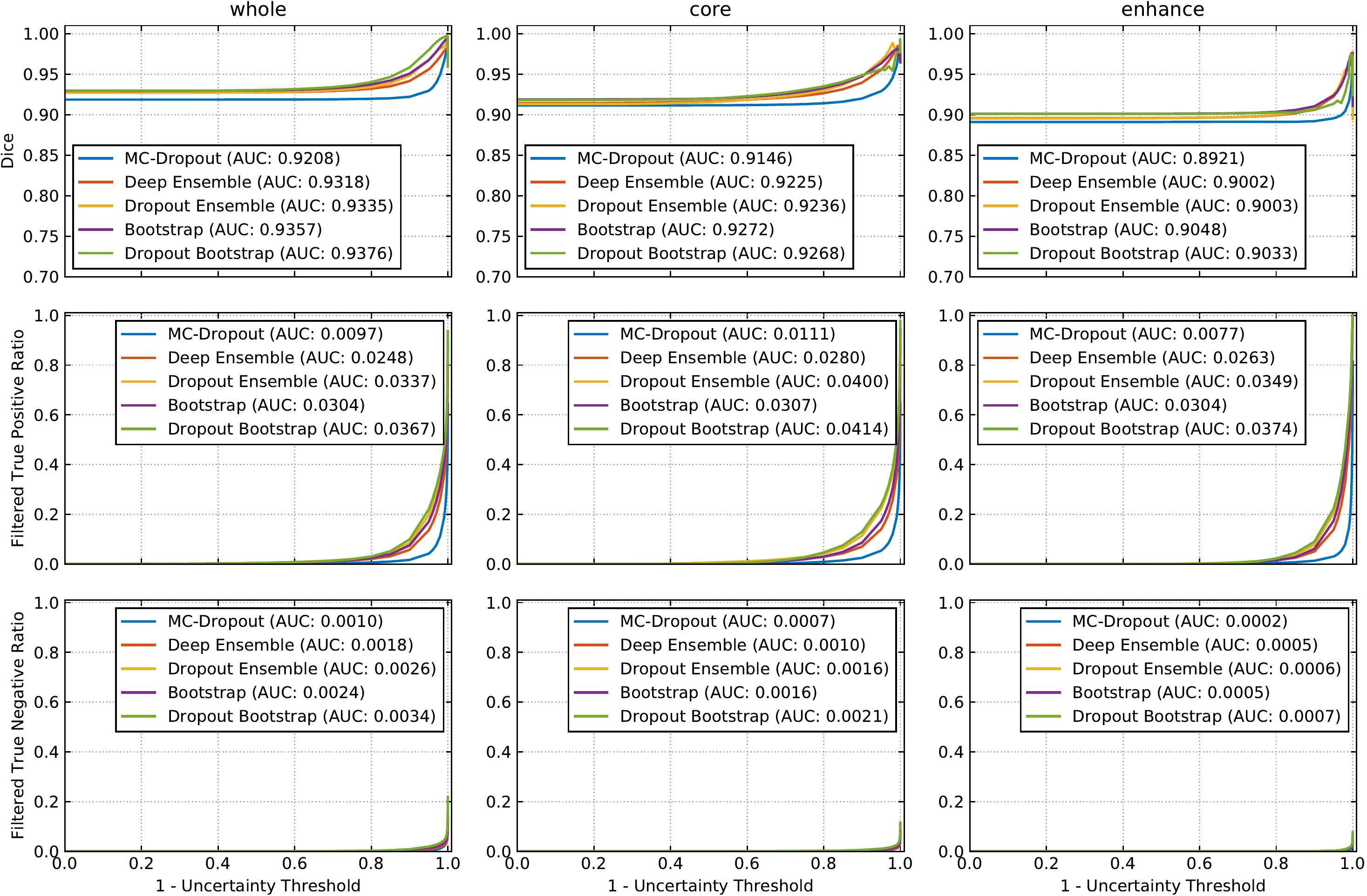}
\center \caption{Effect of changing uncertainty thresholds on 3 different tumour sub-types (whole tumour, core tumour, and enhance tumour) for the MI uncertainty measure.}
\label{fig:UncMI}
\end{figure*}

\begin{table}[h!]
\scriptsize
\centering
\begin{tabular}{|l|c|c|c|}
\hline
\textbf{}                  & \textbf{Whole Tumour} & \textbf{Tumour Core} & \textbf{Enhancing Tumour} \\ \hline
\textbf{MC-Dropout}        & 0.9700                & 0.9676               & 0.9614                    \\ \hline
\textbf{Deep Ensemble}     & 0.9684                & 0.9645               & 0.9578                    \\ \hline
\textbf{Dropout Ensemble}  & 0.9657                & 0.9607               & 0.9549                    \\ \hline
\textbf{Bootstrap}         & 0.9676                & 0.9650               & 0.9580                    \\ \hline
\textbf{Dropout Bootstrap} & 0.9658                & 0.9611               & 0.9551                    \\ \hline
\end{tabular}
\center \caption{An example of the resulting final score (Eq.\ref{Equation-score}) for three different tumour types for the MI uncertainty measure.}
\end{table}

\newpage
%%%%%%%%%%%%%%%%%%%%%%%%%%%%%%%%%%%%%%%%%%%%%%%%%%
\begin{figure*}[h!]
\centering
\includegraphics[width=1.00\textwidth]{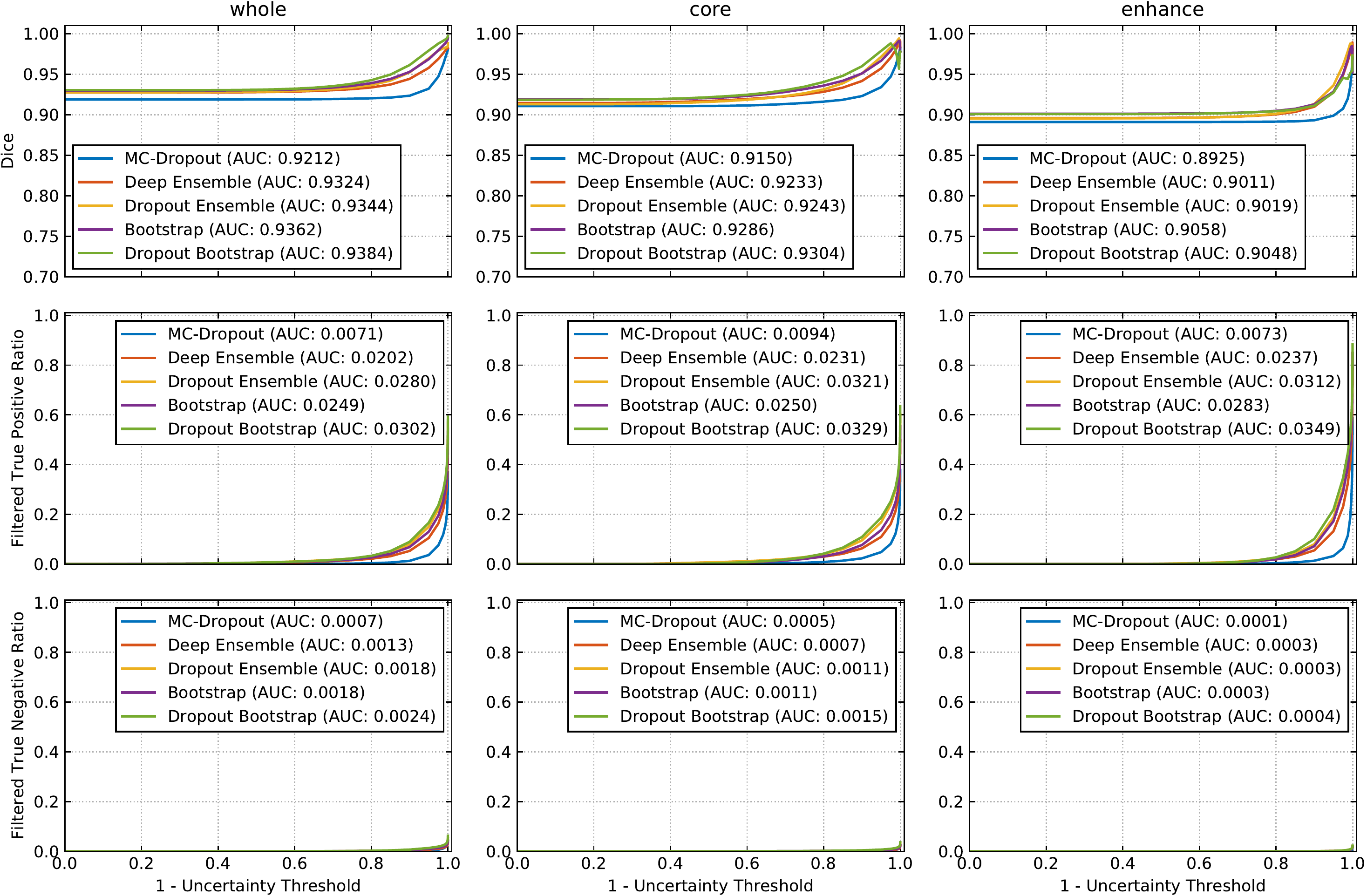}
\center \caption{Effect of changing uncertainty thresholds on 3 different tumour sub-types (whole tumour, core tumour, and enhance tumour) for the sample variance uncertainty measure.}
\label{fig:UncVar}
\end{figure*}

\begin{table}[h!]
\scriptsize
\centering
\begin{tabular}{|l|c|c|c|}
\hline
\textbf{}                  & \textbf{Whole Tumour} & \textbf{Tumour Core} & \textbf{Enhancing Tumour} \\ \hline
\textbf{MC-Dropout}        & 0.9711                & 0.9684               & 0.9617                    \\ \hline
\textbf{Deep Ensemble}     & 0.9703                & 0.9665               & 0.9590                    \\ \hline
\textbf{Dropout Ensemble}  & 0.9682                & 0.9637               & 0.9568                    \\ \hline
\textbf{Bootstrap}         & 0.9698                & 0.9675               & 0.9591                    \\ \hline
\textbf{Dropout Bootstrap} & 0.9686                & 0.9653               & 0.9565                    \\ \hline
\end{tabular}
\center \caption{An example of the resulting final score (Eq.\ref{Equation-score}) for three different tumour types for the sample variance uncertainty measure.}
\end{table}

\newpage

\begin{figure*}[h!]
\centering
\includegraphics[width=1.00\textwidth]{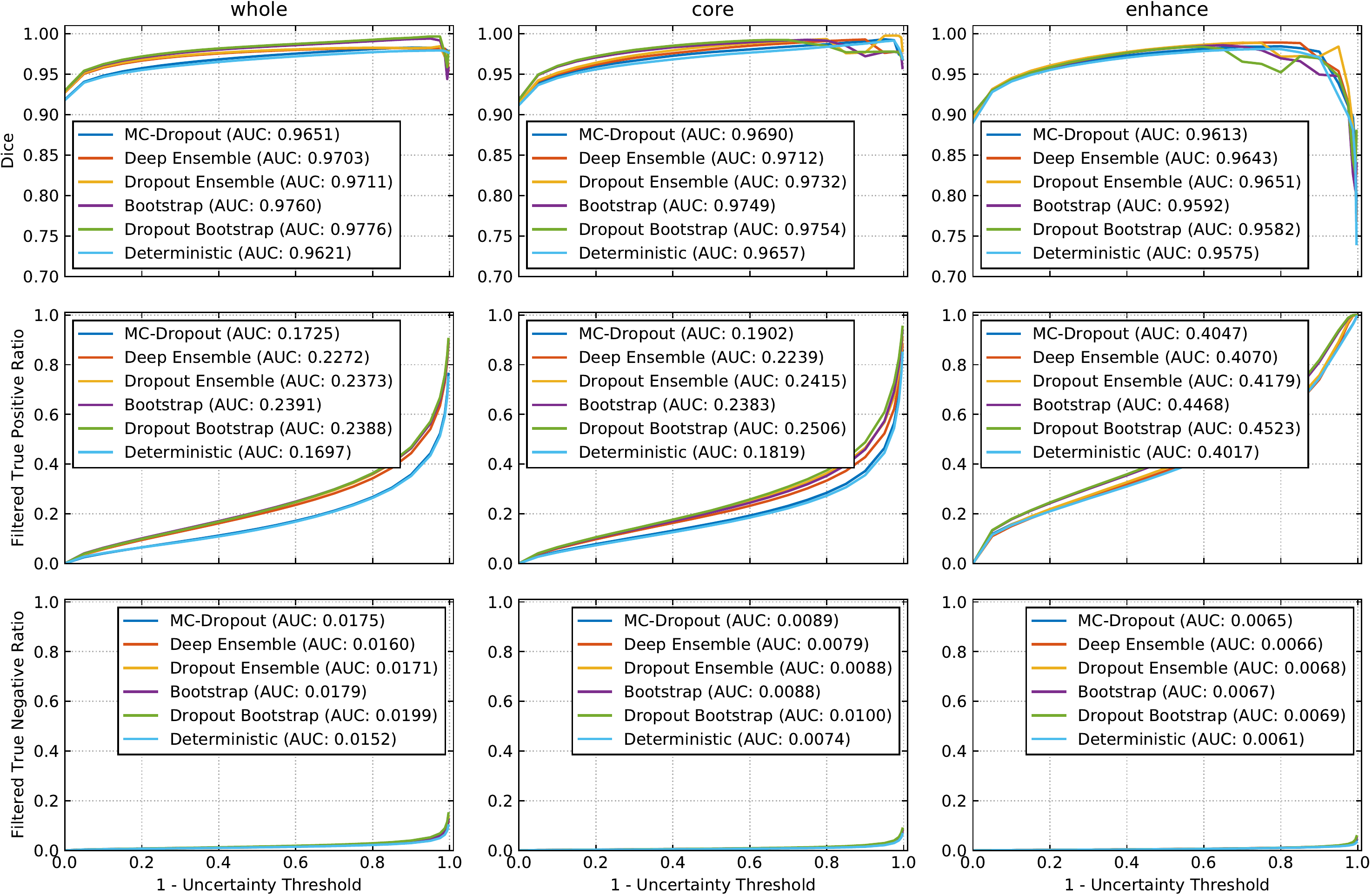}
\center \caption{Effect of changing uncertainty thresholds on 3 different tumour sub-types (whole tumour, core tumour, and enhance tumour) for the entropy uncertainty measure.}
\label{fig:UncEnt}
\end{figure*}

\begin{table}[h!]
\scriptsize
\centering
\begin{tabular}{|l|c|c|c|}
\hline
\textbf{}                  & \textbf{Whole Tumour} & \textbf{Tumour Core} & \textbf{Enhancing Tumour} \\ \hline
\textbf{MC-Dropout}        & 0.9250                & 0.9233               & 0.8500                    \\ \hline
\textbf{Deep Ensemble}     & 0.9090                & 0.9131               & 0.8502                    \\ \hline
\textbf{Dropout Ensemble}  & 0.9056                & 0.9076               & 0.8468                    \\ \hline
\textbf{Bootstrap}         & 0.9063                & 0.9093               & 0.8352                    \\ \hline
\textbf{Dropout Bootstrap} & 0.9063                & 0.9049               & 0.8330                    \\ \hline
\textbf{Determinstic}      & 0.9257                & 0.9255               & 0.8499                    \\ \hline
\end{tabular}
\center \caption{An example of the resulting final score (Eq.\ref{Equation-score}) for three different tumour types for the entropy uncertainty measure.}
\end{table}

\end{document}